\documentclass[pra,aps,showpacs,amsmath,amssymb]{revtex4-1}
\usepackage{physics}
\usepackage{amsmath}
\usepackage{mathrsfs}
\usepackage[font=small,labelfont=bf]{caption}
\usepackage{graphicx}% Include figure files
\usepackage{dcolumn}% Align table columns on decimal point
\usepackage{bm}% bold math
\usepackage{bbm}
\usepackage{bigints}
\begin{document}

%\preprint{APS/123-QED}

\title{Physical and mathematical properties of the space-time-symmetric formalism}% Force line breaks with \\
%\thanks{A footnote to the article title}%

\author{Ricardo Ximenes}%
  \email{ricardoximenes@df.ufpe.br}
 %\altaffiliation[Also at ]{Physics Department, XYZ University.}%Lines break automatically or can be forced with \\
%\author{Fernando Parisio}%
 %\email{parisio@df.ufpe.br}
\author{Eduardo O. Dias}%
 \email{eduardodias@df.ufpe.br}
\affiliation{%
Departamento de Fisica, Universidade Federal de Pernambuco, Recife, Pernambuco 50670-901, Brazil
}%
\date{\today}% It is always \today, today,
             %  but any date may be explicitly specified

\begin{abstract}
It is well known that nonrelativistic quantum mechanics presents a clear asymmetry between space and time. Much of this asymmetry is attributed to the lack of Lorentz invariance of the theory. Nonetheless, a recent work [Phys. Rev. A \textbf{95}, 032133 (2017)] showed that even though this is partially true, there is a broader physical scenario in which space and time can be handled in nonrelativistic quantum theory in a more symmetric way. In this space-time-symmetric formalism, an additional Hilbert space is defined so that time is raised to the status of operator and position becomes a parameter. As a consequence, the Hilbert space now requires a space-conditional quantum state governed by a new quantum dynamics. In this manuscript, we reveal some physical and mathematical properties of the space-time-symmetric formalism such as: symmetries between the Hamilton-Jacobi and the space-conditional equation; the general solution for a time-independent potential; and a new Lagrangian for a spinless particle in one dimensional. Finally, we present the space-conditional equation for a particle under the effect of an electromagnetic field, and the gauge invariance of this equation is proved.
\end{abstract}

\pacs{Valid PACS appear here}% PACS, the Physics and Astronomy
                             % Classification Scheme.
%\keywords{Suggested keywords}%Use showkeys class option if keyword
                              %display desired

\maketitle
%\tableofcontents

\section{\label{sec:level1}Introduction}
The debate about time measurements in quantum mechanics (QM) consists of appealing and controversial issues. One of the main reasons is that time is simply a parameter in traditional QM, whereas the position, for instance, is an operator so that its theoretical predictions are fundamentally probabilistic. In this context, one can interpret $|\psi(x,t)|^2$ as the probability density of finding the particle at the position $x$ \emph{given that} the measurement takes place at the instant $t$. Therefore, besides the fact that time is a parameter, we also verify its conditional character in Schr\"odinger state $\ket{\psi(t)}$. This space-time asymmetric behaviour entails several difficulties in defining a time operator $\hat{T}$ within the sphere of conventional QM. Nonetheless, numerous approaches to address this problem have been proposed in different contexts~\cite{wigner,kijowski,muga,werner,hartle,butkkier,hauge,grot,kumar,giannitrapani,torres,giovanneti,rovelli,halliwell,sels,galapon,spq,timeinqm,timeinqm2,yearsley,reisenberger,marolf,oppenheim,wootters,oldspq}. However, the vast majority of these models had to abdicate either the time-energy canonical relation, $[{\hat T},{\hat H}]=-i\hbar$, or the hermiticity of the time operator. In this scenario, for instance, the problem to calculate a probability distribution for the time that a particle takes to arrive at a spatially localized apparatus is considered by several authors to lay outside the scope of QM. Many other studies have faced similar difficulties, including the tunneling time problem~\cite{butkkier,hauge} and lifetime of metastable systems.

In this puzzling context, the recent works of Ref.~\cite{oldspq,spq} suggest a very different way to address the problem of time in QM. The authors elaborate a space-time-symmetric (STS) formalism where one of the goals is to formulate a time operator in an additional Hilbert space so that both hermiticity and the canonical relation with an energy observable are valid. The STS formalism defines an enlarged Hilbert space $\mathcal{H}=\mathcal{H}_{X} \otimes \mathcal{H}_{T}$ for describing a spinless particle in one dimensional, with $\mathcal{H}_{X}$ being the Hilbert space of traditional QM, and $\mathcal{H}_{T}$ being a new vector space in which the time operator can be safely defined. In addition, Ref.~\cite{spq} defines a space-conditional (SC) quantum state $\ket{\phi(x)}$ belonging to $\mathcal{H}_{T}$ that contains new information not present in the Schr\"odinger state $\ket{\psi(t)}$. Finally, Ref.~\cite{spq} proposes a SC equation of motion that describes how $\ket{\phi(x)}$ evolves \emph{in space}. Recently, the SC formalism has shown to be a promising approach to tackle the tunneling time problem, as can be verified in Ref.~\cite{travt}.

The initial motivation of Ref.~\cite{spq} to formulate the STS formalism is to be able to deal with more general experimental situations where the statistical data is described by a probability distribution $P(x,t)$ of finding the particle in a given region of space $[x,x+dx]$ and in a certain interval of time $[t,t+dt]$. According to Bayes rule, the joint probability distribution $P(x,t)$ is equal to the probability density of finding the particle at the position $x$ \emph{given that} the measurement occurs precisely at $t$, $|\psi(x,t)|^2$, times the probability density of the system being measured at the instant $t$, $f(t)$, whatever the outcome. In this broader scenario, we have
\begin{equation}
\label{P1}
P(x,t)dxdt=|\psi(x|t)|^2f(t)~dxdt.
\end{equation}
From now on, due to the time conditional character of the Schr\"odinger wave function, we use the more appropriate notation $\psi(x|t)$. It is important to understand that the function $f(t)$ cannot be obtained exclusively through the Sch\"odinger state $|\psi(t)\rangle$. The distributions $f(t)$ and $P(x,t)$ correspond to new probability densities that are essential to describe statistical data in which position and time are measured simultaneously. Finally, due to the symmetry of Bayes rule, $P(x,t)$ can be written as
\begin{equation}
\label{P2}
P(x,t)dxdt=|\phi(t|x)|^2g(x)~dxdt.
\end{equation}
In Eq.~(\ref{P2}) note that $x$ and $t$ play opposite roles in comparison with $x$ and $t$ in Eq.~(\ref{P1}), which includes the Schr\"odinger state $\psi(x|t)$. Here $t$ must be seen as the eigenvalue of a temporal hermitian observable that will be defined later, and $x$ is a continuous parameter that we can choose with arbitrary precision in order to evaluate the time dependence of $\phi(t|x)$. In these circumstances, $|\phi(t|x)|^2dt$ represents the probability of finding the particle in the time interval $[t,t+dt]$ \emph{given that} the measurement takes place at the position $x$, and $g(x)$ is the probability distribution of position measurements irrespective of the time at which they happen. Notice that in this broader picture, these new probability distributions are outside the scope of traditional QM.

In this new framework, the parameter $t$ becomes the eigenvalue of the operator $\hat{T}$ that represents the time at which the system is observed. Therefore, one can observe the symmetry between $\hat{X}|x\rangle=x|x\rangle$ in Sch\"odinger QM and $\hat{T}|t\rangle=t|t\rangle$ in the SC formalism. The SC wave function $\phi(t|x)$ is obtained similarly to traditional QM [$\langle x|\psi(t)\rangle=\psi(x|t)$] by the calculation $\langle t|\phi(x)\rangle=\phi(t|x)$. With these facts in mind, one can verify that temporal predictions should be made by states belonging to $\mathscr{H}_{T}$ instead of $\mathscr{H}_{X}$. In this regard, the SC equation has to be solved, and as we will see in the next sections, it is quite difficult to deal with it since this equation encompasses the square root of operators. However, for time-independent potentials, which include a wide range of physical situations, we will show that this problem can be overcome in a simple manner.

In this manuscript, we further develop the STS formalism by focusing on the new wave function $\phi(t|x)$ as follows: in Sec.~\ref{sec:level2} we reveal physical symmetries between the Hamilton-Jacobi equation, the space-conditional QM, and Schr\"odinger QM. Then, in Sec~\ref{sec:level3} a brief summary of the results elaborated in Ref.~\cite{spq} is presented. In Sec~\ref{sec:level4} the Lagrangian for the SC equation is obtained, and in Sec.~\ref{sec:level5} we solve $\phi(t|x)$ for a general time-independent potential $V(x)$. Finally, in Sec.~\ref{sec:level6} we present the SC equation for a particle subjected to an electromagnetic field, and then the gauge invariance of the STS formalism is proved.

\section{\label{sec:level}Mathematical and physical properties of the Space-conditional equation}

\subsection{\label{sec:level2}Symmetries between quantum and classical physics}
The action $S$ is a quantity of utmost importance in physics. It plays a major role, not only in classical mechanics, but also in theoretical physics in general. In this section, differently from the approach of Ref.~\cite{spq}, we derive the STS formalism by using the action as the starting point. More specifically, we show that the Hamilton-Jacobi equation, together with Sch\"odinger QM, will provide symmetry arguments in order to formulate the SC equation naturally. First, recall that in classical mechanics the Hamilton-Jacobi equation reads
\begin{equation}\label{hj}
H\left(x,p;t\right)=-\pdv{S}{t}\left(x,p;t\right) ~~~{\rm and}~~~ p=\pdv{S}{x}\left(x,p;t\right) .
\end{equation}
Here, we use the notation with $t$ on the right side of the parentheses to indicate that it is just a parameter, in contrast to $x$, which is a function to be calculated. In order to obtain traditional QM from the Hamilton-Jacobi equation, let us follow a similar approach to the one used by Schr\"odinger in Ref.~\cite{sco}. Suppose that the action is the phase of some physical process described by a wave function $\psi=e^{iS/\hbar}$. By isolating $S$ in this relation, $S=-i\hbar\ln{\psi}$, and substituting it back into the Hamilton-Jacobi equation, we have
\begin{equation}\label{schro1}
H\left(x,p;t\right)\psi=i\hbar\pdv{\psi}{t}~~~{\rm and}~~~ p~\psi=-i\hbar\pdv{\psi}{x}.
\end{equation}
By inspecting the expression on the right of Eq.~(\ref{schro1}), we automatically identify the momentum $p$ as the operator ${\hat p}=-ihd/dx$, and the Hamiltonian as the operator ${\hat H}={\hat H}(x,{\hat p};t)$. Note that this kind of quantization keeps $t$ as a parameter, and transforms the Hamilton-Jacobi equation into the Schr\"odinger equation
\begin{equation}\label{scheq}
\hat{H}(x,\hat{p};t)~\psi(x,t)=i\hbar\pdv{\psi}{t} (x,t)~~~{\rm and}~~~ \hat{p}~\psi(x,t)=-i\hbar\pdv{\psi}{x} (x,t).
\end{equation}
This context is extremely unfavorable to define any kind of time operator since time is considered as a parameter from the beginning of the physical description. As a result, the following idea arises naturally: if we switch the roles of space and time in the Hamilton-Jacobi equation, that is, if space becomes a parameter and time a function to be calculated, would it be possible to formulate a quantum theory where a time operator is a natural result? To answer this question, first, for the sake of convenience, let us define $h\equiv H$, and write $p$ in terms of $h$ and $x$ so that $p=P(t,h;x)=\pm[2m(h-V(x))]^{1/2}$. Under these circumstances, we have
\begin{equation}\label{hj2}
h=-\pdv{S}{t}\left(t,h;x\right) ~~~{\rm and}~~~ P\left(t,h;x\right)=\pdv{S}{x}\left(t,h;x\right).
\end{equation}
Similarly to the previous calculation, by supposing that the action $S$ is the phase of a physical process so that $\phi=e^{iS/\hbar}$, we have
\begin{equation}\label{hh}
h~\phi=i\hbar\pdv{\phi}{t}~~~{\rm and}~~~ P\left(t,h;x\right)\phi=-i\hbar\pdv{\phi}{x}.
\end{equation}
It is worth to notice the symmetry between Eqs.~(\ref{schro1}) and (\ref{hh}). Now, in contrast to the procedure to obtain the Schr\"odinger equation, we identify $h$ as the operator ${\hat h}=ihd/dt$, and the momentum ${\hat P}={\hat P}(x,h;t)$ as the operator responsible for the \emph{space} evolution of wave function $\phi(t,x)$. Therefore, by writing in a more appropriate notation, we have
\begin{equation}\label{schmieq}
\hat{h}~\phi(t,x)=i\hbar\pdv{\phi}{t} (t,x)~~~{\rm and}~~~\hat{P}(\hat{t},\hat{h};x)~\phi(t,x)=-i\hbar\pdv{\phi}{x} (t,x),
\end{equation}
which is exactly the space-conditional equation for $\phi(t|x)$ derived in Ref.~\cite{spq}. Note that now the dynamic equation involves a space derivative instead of the time derivative observed in the Schr\"odinger equation. This opposite behaviour induces us to treat $x$ as a dynamic variable and $t$ as a parameter.

In this new quantum formalism, a time operator $\hat T$ such that ${\hat T}|t\rangle=t|t\rangle$, and a new state $|\phi(x)\rangle$, such that $\phi(t|x)=\langle t|\phi(x)\rangle$, are defined. Finally, it is interesting to notice that, even tough this two formalism describe different physical situations, the fact that both results \eqref{scheq} and \eqref{schmieq} emerge from the Hamilton-Jacobi equation means that, in the classical limit, there is no difference in treating either $x$ or $t$ as a parameter. After all, this result is expected since the conditional character of time or position is irrelevant in classical physics.

Before moving on to the next section, one may notice that we can go from \eqref{scheq} to \eqref{schmieq}, or vice-versa, via the transformation $(x,P)  \Leftrightarrow (t,-H)$. This is an important relation that will be recurrent throughout this work.

\subsection{\label{sec:level3}Summary of the space-time-symmetry formalism}
In this section, we present a summary of the STS formalism as it was formulated in Ref.~\cite{spq}. First, let us state that a complete description of a non-relativistic quantum particle requires an enlarged Hilbert space $\mathscr{H}_{X} \otimes \mathscr{H}_{T}$. Recall that $\mathscr{H}_{X}$ consists of the Hilbert space of the usual QM, whereas $\mathscr{H}_{T}$ is an extension of the vector space of QM where the roles of space and time are switched. Here, the notation of the standard observables of the traditional QM, which act in $\mathscr{H}_{X}$, are denoted by: $\hat{X}$ (position operator), $\hat{p}$ (momentum operator) and $\hat{H}(\hat{X},\hat{p};t)$ (Hamiltonian operator). To each of these observables, there is an analogous operator acting in the ``mirror'' space $\mathcal{H}_{T}$. These operators are: the time operator $\hat{T}$ (mirror of $\hat{X}$), Hamiltonian $\hat{h}$ (mirror of $\hat{p}$), and momentum $\hat{P}(\hat{T},\hat{p};x)$ (mirror of $\hat{H}$). In this context, it is natural to call $\hat h$ the energy operator, and $\hat P$ the ``\emph{Pamiltonian}'' operator.

In QM whose states belong to the Hilbert Space $\mathscr{H}_{X}$, time is a parameter, whereas the position is an operator so that $\hat{X}\ket{x}=x\ket{x}$. Additionally, this operator is required to satisfy the canonical commutation relation $[\hat{X},\hat{p}]=i\hbar$. All the physical information in this space is contained in the state ket ${\ket{\psi(t)}}$, which obeys the Schr\"odinger equation
\begin{equation}\label{sc}
\hat{H}\ket{\psi(t)}=i\hbar \frac{d}{dt}\ket{\psi(t)}.
\end{equation}
The Hamiltonian operator $\hat{H}$ can be obtained from the classical Hamiltonian of the system via the usual quantization rule
\begin{equation}\label{qntrule1}
H(x,p;t)=\frac{p^2}{2m}+V(x,t) ~~ \Rightarrow ~~ \hat{H}(\hat{X},\hat{p};t)=\frac{\hat{p}^2}{2m}+V(\hat{X},t).
\end{equation}
By substituting the explicit form of the Hamiltonian operator into \eqref{qntrule1}, we have
\begin{equation}\label{schro}
\left[\frac{\hat{p}^2}{2m}+V(\hat{X},t)\right]\ket{\psi(t)}=i\hbar \frac{d}{dt}\ket{\psi(t)},
\end{equation}
Here, the momentum operator is defined as $\mel{x}{\hat{p}}{\psi(t)}=-i\hbar\partial_{x}\psi(x|t)$ so that in the position representation Eq.~\ref{schro} reads
\begin{equation}
\left[-\frac{\hbar^2}{2m}\pdv[2]{x}+V(x,t)\right] \psi(x|t)=i\hbar \pdv{\psi}{t} (x|t)
\end{equation}
Finally, the probability density is
\begin{equation}\label{rho}
\rho(x|t) \dd{x}=\psi^{*}(x|t)\psi(x|t) \dd{x}.
\end{equation}
Eq.~(\ref{rho}) represents the probability of a position measurement having an outcome in the interval $[x,x+\dd{x}]$, \textit{given that} the measurement occurred at the time t. The interpretation of this time conditional probability density is the main difference between traditional QM and space-conditional QM. In this context, $\rho(x|t)$ describes experimental outcomes when the position statistical data are evaluated for a fixed instant of time.

On the other hand, in QM described in the Hilbert space $\mathscr{H}_{T}$, position plays the role of parameter, and time behaves as an operator. Moreover, analogous to $[\hat{X},\hat{p}]=i\hbar$, the time operator $\hat T$ satisfies the canonical commutation relation $[\hat{T},\hat{h}]=-i\hbar$. Now, the physical information in this space is contained in the new ket $\ket{\phi(x)}$, which obeys the SC equation
\begin{equation}\label{diaseqvec}
\hat{P}\ket{\phi(x)}=-i\hbar \frac{d}{dx}\ket{\phi(x)}.
\end{equation}
The Pamiltonian operator $\hat{P}$ can be obtained from the classical momentum of the system via the new quantization rule,
\begin{equation}\label{quantrule}
p(t,H;x)=\pm \sqrt{2m\big [H-V(x,t)\big ]} \quad \Rightarrow  \quad \hat{P}(\hat{T},\hat{h};x)=\hat{\sigma}_{z} \sqrt{2m\left[\hat{h}-V(x,\hat{T})\right]}.
\end{equation}
By the substitution  of the Pamiltonian operator~\eqref{quantrule} back into \eqref{diaseqvec}, we obtain
\begin{equation}\label{diaseq}
\hat{\sigma}_{z} \sqrt{2m\left[\hat{h}-V(x,\hat{T})\right]}\ket{\phi(x)}=-i\hbar \frac{d}{dx}\ket{\phi(x)}.
\end{equation}
Here the Hamiltonian operator $\hat{h}$ is defined as $\mel{t}{\hat{h}}{\phi(x)}=i\hbar\partial_{t}\phi(t|x)$. In these circumstances, Eq.~(\ref{diaseq}) in the time representation $|t\rangle$ reads
\begin{equation}\label{diaseqtrep}
\hat{\sigma}_{z} \sqrt{2m\left(i\hbar \pdv{t}-V(x,t)\right)}\phi(t|x)=-i\hbar \pdv{\phi}{x} (t|x),
\end{equation}
where $\phi(t|x)$ is a two-component vector
\begin{equation}
\phi(t|x)=\mqty(\phi^{+}(t|x)  \\  \phi^{-}(t|x)).
\end{equation}
In this way, the probability density is given by
\begin{equation}\label{rhodias}
\rho(t|x)=\phi^{\dagger}(t|x)\phi(t|x).
\end{equation}
Here, the symbol $\dagger$ is the conjugate transpose operator. Notice that the important difference between $\rho_{\phi}$ and $\rho_{\psi}$: now, $\rho_{\phi}(t|x)\dd{t}$ represents the probability of measuring the particle in the time interval $[t,t+\dd{t}]$, \textit{given that the measurement occurred at the position x}. Moreover, it is important to note the similarities between all the equations of Schr\"odinger QM~(\ref{sc})-(\ref{rho}) and the equations of space-conditional QM~(\ref{diaseqvec})-(\ref{rhodias}).

\subsection{\label{sec:level4}Lagrangian of the space-conditional equation}
It is well known that the Schr\"odinger equation can be deduced from the Lagrangian
\begin{eqnarray}\label{lagsch}
\mathcal{L}_{\psi}(\psi,\psi^{*},\partial_{x}^{2}\psi,\partial_{t}\psi;x,t)&=&\psi^{*}(x|t)\left[\hat{H}-i\hbar\pdv{t}\right]\psi(x|t)\nonumber\\
&=&\psi^{*}(x|t)\left[\frac{\hbar^2}{2m}\pdv[2]{x}-V(x,t)+i\hbar\pdv{t}\right]\psi(x|t)
\end{eqnarray}
By symmetry, we expect that the Lagrangian for the SC wave function $\phi(t|x)$ should be obtained via the transformation $(t,H) \Rightarrow (x,-P)$
\begin{eqnarray}\label{lagsts}
\mathcal{L}_{\phi}(\phi,\phi^{\dagger},\partial_{x}\phi,\partial^{1/2}_{t}\phi;x,t)&=&\phi^{\dagger}(t|x)\left[\hat{P}+i\hbar\pdv{x}\right]\phi(t|x)\nonumber\\
&=&\phi^{\dagger}(t|x)\left[\hat{\sigma}_{z}\sqrt{2m\left(i\hbar \pdv{t}-V(x)\right)}+i\hbar\pdv{x}\right]\phi(t|x).
\end{eqnarray}
In fact, below we show that Eq.~(\ref{lagsts}) is the Lagragian for $\phi(t|x)$ for time-independent potential situations. The general case involves the square root of a time derivative and until now, we have not found out the Lagrangean for this broader scenario. To verify the validity of Eq.~(\ref{lagsts}) for the time-independent case, let us write $\phi(t|x)=e^{-i\varepsilon t/\hbar}\phi_{\varepsilon}(x)$, where $\varepsilon$ is the eigenvalue of the hamiltonian operator, $\hat{h}\ket{\varepsilon}=\varepsilon \ket{\varepsilon}$. By substituting this expression into $\mathcal{L}_{\phi}$, we have
\begin{equation}
\mathcal{L}_{\phi}(\phi,\phi^{\dagger},\partial_{x}\phi;x)=\phi^{\dagger}(t|x)\left[\hat{\sigma}_{z}\sqrt{2m\big[\varepsilon-V(x)\big]}+i\hbar\pdv{x}\right]\phi(t|x).
\end{equation}
Now, by applying the Euler-Lagrange equation to the field $\phi(t|x)$,
\begin{equation}
\pdv{x}\left(\pdv{\mathcal{L}_{\phi}}{(\pdv{\phi}{x})} \right) -\pdv{\mathcal{L}_{\phi}}{\phi}=0,
\end{equation}
we obtain
\begin{equation}
i\hbar \pdv{\phi^{\dagger}}{x} (t|x)-\phi^{\dagger}(t|x)\hat{\sigma}_{z}\sqrt{2m\big [\varepsilon-V(x)\big ]}=0.
\end{equation}
By taking the complex conjugate of this equation,
\begin{equation}\label{ip}
-i\hbar \pdv{\phi}{x} (t|x)=\hat{\sigma}_{z}\sqrt{2m\big [\varepsilon-V(x)\big ]}\phi(t|x),
\end{equation}
which is the SC equation~(\ref{diaseqtrep}) for a time-independent potential.

\subsection{\label{sec:level5}General solution of $\phi(t|x)$ for time-independent potentials}
As we discussed in last section, for time-independent potentials, the SC equation does not contain a square root of operators so that it is much easier to solve it. By using $\phi(t|x)=e^{-i\varepsilon t/\hbar}\phi_{\varepsilon}(x)$, Eq.~(\ref{ip}) becomes
\begin{equation}
\hat{\sigma}_{z}\sqrt{2m\big[\varepsilon-V(x)\big]}\phi_{\varepsilon}(x)=-i\hbar\frac {d}{dx}{\phi_{\varepsilon}}(x).
\end{equation}
This first order differential equation has a very simple solution given by
\begin{equation}
\phi_{\varepsilon}(x)=\frac{1}{\sqrt{2\pi \hbar}}e^{(i/\hbar)\hat{\sigma}_{z}\bigintsss\sqrt{2m\big [\varepsilon-V(x')\big ]}\dd{x'}}.
\end{equation}
Therefore, the general solution can be written as
\begin{equation}\label{phig}
\phi(t|x)=\frac{1}{\sqrt{2\pi \hbar}}\int_{-\infty}^{\infty} e^{(i/\hbar)\hat{\sigma}_{z}\bigintsss \sqrt{2m\big [\varepsilon-V(x')\big ]}\dd{x'}-i\varepsilon t/\hbar} ~ C_{\varepsilon}\dd{\varepsilon},
\end{equation}
We point out that $C_{\varepsilon}$ is a vector of two components.

It is important to address the interesting similarity between Eq.~(\ref{phig}), which is valid for potentials such as $V=V(x)$, and the solution of Sch\"oridinger equation for potentials that depend exclusively on time $V(x,t)=V(t)$, which is given by
\begin{equation}\label{psig}
\psi(x|t)=\frac{1}{\sqrt{2\pi \hbar}}\int_{-\infty}^{\infty} e^{-(i/\hbar)\bigintsss H(p;t') \dd{t'}+ipx/\hbar}~A_{p}\dd{p}.
\end{equation}
Note the Hamiltonian generating the time evolution of $\psi(x|t)$ in Eq~(\ref{psig}), whereas the Pamiltonian performing the \emph{space} evolution of $\phi(t|x)$ in Eq.~(\ref{phig}). Here again we observe that one can readily go from one solution to the other via the transformation $(t,H) \Rightarrow (x,-P)$. This symmetry can become even more apparent if we identify the classical action $S$ in each case,
\begin{equation}
\phi(t|x)=\frac{1}{\sqrt{2\pi \hbar}}\int_{-\infty}^{\infty} \dd{\varepsilon} e^{-(i/\hbar)S(\varepsilon,x)}C_{\varepsilon}, \quad \quad V(x,t)=V(x).
\end{equation}
\begin{equation}
\psi(x|t)=\frac{1}{\sqrt{2\pi \hbar}}\int_{-\infty}^{\infty}\dd{p} e^{-(i/\hbar)S(p;t)}A_{p}, \quad \quad V(x,t)=V(t),
\end{equation}

\subsection{\label{sec:level6}The space-conditional equation under the influence of an electromagnetic field and the gauge invariance}
One important symmetry that is required in any quantum formalism, in our case an extension of the Schr\"odinger description of QM, that takes into account electromagnetic forces must be the gauge invariance. This means that regardless of the gauge choice for the electromagnetic potentials, the probability density and the SC equation must remain unchanged. In this section we verify the gauge invariance of the STS formalism, which gives more support to the validity of the theory. The Hamiltonian in the presence of an electromagnetic field is given by
\begin{equation}\label{hamil}
H=\frac{1}{2m}\left(p-\frac{q}{c}A(x,t)\right)^2+q\Phi(x,t).
\end{equation}
In order to construct the Pamiltonian let us isolate the kinematic momentum in Eq.~(\ref{hamil}),
\begin{equation}
p=\pm\sqrt{2m\big [H-q\Phi(x,t)\big ]}+\frac{q}{c}A(x,t).
\end{equation}
By using the quantisation rule of Eq.\eqref{quantrule},
\begin{equation}\label{pamel}
\hat{P}({\hat T},{\hat h})=\hat{\sigma}_{z}\sqrt{2m\left [\hat{h}-q\hat{\Phi}(x,\hat{T})\right ]}+\frac{q}{c}\hat{A}(x,\hat{T}).
\end{equation}
By the substitution of the Pamiltonian~(\ref{pamel}) in the SC equation~(\ref{diaseqtrep}), we have
\begin{equation}
\left[\hat{\sigma}_{z}\sqrt{2m\left [\hat{h}-q\hat{\Phi}(x,\hat{T})\right ]}+\frac{q}{c}\hat{A}(x,\hat{T})\right]\ket{\phi(x)} =-i\hbar\frac{d}{dx} \ket{\phi(x)},
\end{equation}
and, in the time representation $|t\rangle$, this equation becomes
\begin{equation}\label{oldcalib}
\left[\hat{\sigma}_{z}\sqrt{2m\left (i\hbar\pdv{t}-q\Phi(x,t)\right)}+\frac{q}{c}A(x,t)\right]\phi(t|x) =-i\hbar \pdv{x}\phi(x|t).
\end{equation}

The gauge transformation for the potentials $\Phi(x,t)$ and $A(x,t)$ is given by
\begin{equation}\label{gaugechange}
\Phi'(x,t)=\Phi-\frac{1}{c}\pdv{F}{t}(x,t), \quad A'(x,t)=A(x,t)+\pdv{F}{x}(x,t).
\end{equation}
In our current case, the SC formalism will be gauge invariant if the transformed SC equation
\begin{equation}\label{newphi}
\left[\hat{\sigma}_{z}\sqrt{2m\left (i\hbar\pdv{t}-q\Phi'(x,t)\right )}+\frac{q}{c}A'(x,t)\right]\phi'(t|x) = -i\hbar \pdv{x}\phi'(t|x)
\end{equation}
results in the original SC equation~(\ref{oldcalib}) when the new wave function $\phi'(t|x)$ is related to the original one $\phi(t|x)$ via a unitary transformation $\hat{U}$, i.e. $\phi'(t|x)=\hat{U}\phi(t|x)$. In these circumstances,
\begin{equation}\label{unitransform}
\rho'(t|x)=\phi'^{\dagger}(t|x)\phi'(t|x)=\phi^{\dagger}(t|x)\phi(t|x)=\rho(t|x),
\end{equation}
that is, the probability density is independent of the gauge choice. To prove this statement, first, for convenience let us make the most simple choice for the unitary transformation given by $\hat{U}=\exp{iqF(x,t)/(\hbar c)}$. By substituting $\phi'(x,t)$ by $e^{\frac{iq}{\hbar c}F(x,t)} \phi(x,t)$ in Eq.~(\ref{newphi}), we have
\begin{equation}\label{newcali}
\hat{\sigma}_{z}\sqrt{2m\left(i\hbar\pdv{t}-q\Phi'(x,t)\right)}e^{\frac{iq}{\hbar c}F(x,t)} \phi(t|x)+\frac{q}{c} \left( A'(x,t)-\pdv{x} F(x,t)\right) e^{\frac{iq}{\hbar c}F(x,t)} \phi(t|x)= -i\hbar ~ e^{\frac{iq}{\hbar c}F(x,t)} \pdv{x} \phi(t|x)).
\end{equation}
To calculate the square root on the left side of Eq.~(\ref{newcali}), let us write it as a power series:
\begin{equation}\label{exp}
\sqrt{i\hbar\pdv{t}-q\Phi'(x,t)}=\sum_{n=0}^{\infty}c_{n}{\left(i\hbar\pdv{t}-q\Phi(x,t)\right)^{n}}.
\end{equation}
In order to visualize the resulting expression after the application of the operator~(\ref{exp}) on $\phi'(x,t)$, first notice that for $n=1$ we have
\begin{equation}\label{prop1}
\left(i\hbar\pdv{t}-q\Phi'(x,t)\right) e^{\frac{iq}{\hbar c}F(x,t)}\phi(t|x) =e^{\frac{iq}{\hbar c}F(x,t)}\left [i\hbar\pdv{t}-q\left(\Phi'(x,t)+1/c \pdv{F}{t}\right )\right ] \phi(t|x).
\end{equation}
By inspecting Eq.~(\ref{prop1}), we notice that for $n=2$ the second application results in
\begin{equation}
\left(i\hbar\pdv{t}-q\Phi'(x,t)\right)^2 e^{\frac{iq}{\hbar c}F(x,t)}\phi(t|x)=e^{\frac{iq}{\hbar c}F(x,t)}\left[i\hbar\pdv{t}-q\left(\Phi'(x,t)+1/c \pdv{F}{t}\right)\right]^2 \phi(t|x).
\end{equation}
Therefore, applying $n$ times we readily visualize that
\begin{equation}
{\left(i\hbar\pdv{t}-q\Phi'(x,t)\right)^{n}}e^{\frac{iq}{\hbar c}F(x,t)} \phi(t|x)=e^{\frac{iq}{\hbar c}F(x,t)}{\left[i\hbar\pdv{t}-q\left(\Phi'(x,t)+1/c \pdv{F}{t}\right)\right]^{n}} \phi(t|x).
\end{equation}
As a consequence, the square root operator in~\eqref{exp} when applied to $\phi'(x,t)$ can be written as
\begin{equation}
\sqrt{i\hbar\pdv{t}-q\Phi'(x,t)}e^{\frac{iq}{\hbar c}F(x,t)} \phi(x,t)=e^{\frac{iq}{\hbar c}F(x,t)}\sqrt{i\hbar\pdv{t}-q\left(\Phi'(x,t)+1/c \pdv{f}{t}\right)} \phi(x,t).
\end{equation}
By substituting this result back into \eqref{newcali}, we obtain
\begin{eqnarray}\label{newcali2}
e^{\frac{iq}{\hbar c}F(x,t)}\hat{\sigma}_{z}\sqrt{2m\left[i\hbar\pdv{t}-q\left(\Phi'(x,t)+1/c \pdv{F}{t}\right)\right]} \phi(t|x)+\frac{q}{c} \left( A'(x,t)-\pdv{x} F(x,t)\right) e^{\frac{iq}{\hbar c}F(x,t)} \phi(t|x)= \nonumber\\ = -i\hbar e^{\frac{iq}{\hbar c}F(x,t)} \pdv{x} \phi(t|x).
\end{eqnarray}
Finally, by cancelling the exponential factor in both sides of Eq.~(\ref{newcali2}), and using the gauge transformations~\eqref{gaugechange}, we have
\begin{equation}\label{newcali3}
\hat{\sigma}_{z}\sqrt{2m \left(i\hbar\pdv{t}-q\Phi(x,t)\right)} ~ \phi(t|x)+\frac{q}{c} A(x,t) \phi(t|x)= -i\hbar \pdv{x} \phi(t|x).
\end{equation}
Eq.~(\ref{newcali3}) concludes our proof that the SC equation is gauge invariant.

\section{\label{sec:level7}Conclusion}
The asymmetry between space and time in traditional QM has been mostly attributed to the lack of Lorentz covariance of the theory. Therefore, it is commonly argued that the only way to remove this asymmetric nature is exclusively through a relativistic approach. Nevertheless, the results of the recent work of Ref.~\cite{spq}, which formulates a STS theory that is further developed here, show that this argument is not completely true. In fact, it is possible to construct a quantum theory, without requiring Lorentz covariance, in which space and time develops a much more symmetric role.

The STS formalism proposes that the traditional nonrelativistic QM is part of a broader theory in which an additional Hilbert space is composed of SC states $|\phi(x)\rangle$ where position is a parameter and time is an operator. In this SC description, the dynamic of $|\phi(x)\rangle$ is dictated by a equation that governs its \emph{space} evolution. Here we present the quantization rules for the STS formalism by using the Hamilton-Jacobi equation, as well as their symmetries with the Sch\"odinger quantum theory. Moreover, a relevant connection between the Hamilton-Jacobi formalism and the QM equations, including the space-conditional QM, is revealed. Then we derive the general solution for the SC equation for time-independent potentials, and by inspecting the Sch\"odinger solution for space-independent potentials, we verified that both situations are connected via the transformation $(t,H) \Rightarrow (x,-P)$. In the next section, a Lagrangian for the SC equation is obtained and, as expected, we also linked it with the Lagrangian of the Sch\"odinger equation via $(t,H) \Rightarrow (x,-P)$.  Finally, we present the SC equation for a particle under the effect of the electromagnetic field, and the gauge invariance of the STS formalism is proved, as it should be.

%The STS formalism had its preliminary ideas published first in Ref. Only later in Ref.~\cite{spq} was this new approach in some way concluded. Apart from the results of this preliminary work, there are several aspects to be addressed and questions to be answered that we intend to continue to investigate.

\section{\label{sec:level8}acknowledgements}
The authors thank CNPq and FACEPE (Brazilian Research
Councils) for financial support. E. O. D. acknowledges
financial support from FACEPE through PPP
Project No. APQ-0800-1.05/14.


\newpage
\begin{thebibliography}{9}
\bibitem{wigner}
E. P. Wigner, Phys. Rev. 98, 145 (1955).
\bibitem{kijowski}
J. Kijowski, Rep. Math. Phys. 6, 361 (1974).
\bibitem{muga}
J. S. Muga, S. Brouard, and D. Macas, Ann. Phys. (NY) 240,
351 (1995).
\bibitem{werner}
R. Werner, J. Math. Phys. 27, 793 (1986).
\bibitem{hartle}
J. B. Hartle, in Gravitation and Quantizations, Proceedings of
the 1992 Les Houches Summer School, edited by B. Julia and
J. Zinn-Justin (North-Holland, Amsterdam, 1994).
\bibitem{grot}
N. Grot, C. Rovelli, and R. S. Tate, Phys. Rev. A 54, 4676 (1996).
\bibitem{kumar}
N. Kumar, Pramana 25, 363 (1985).
\bibitem{giannitrapani}
R. Giannitrapani, Int. J. Theor. Phys. 36, 1575 (1997).
\bibitem{torres}
G. Torres-Vega, Phys. Rev. A 75, 032112 (2007).
\bibitem{giovanneti}
V. Giovannetti, S. Lloyd, and L. Maccone, Phys. Rev. D 92,
045033 (2015).
\bibitem{rovelli}
C. Rovelli, Phys. Rev. D 43, 442 (1991); 42, 2638 (1990).
\bibitem{halliwell}
J. J. Halliwell, J. Evaeus, J. London, and Y. Malik, Phys. Lett.
\bibitem{sels}
D. Sels and M. Wouters, arXiv:1501.05567.
\bibitem{galapon}
E. A. Galapon, R. F. Caballar, and R. T. Bahague, Jr., Phys. Rev.
Lett. 93, 180406 (2004).
\bibitem{timeinqm}
Time in Quantum Mechanics, edited by J. G. Muga, R. Sala
Mayato, and I. L. Egusquiza (Springer, Berlin, 2002).
\bibitem{timeinqm2}
Time in Quantum Mechanics, edited by J. G. Muga, A.
Ruschhaupt, and A. del Campo (Springer, Berlin, 2009),
Vol. 2.
\bibitem{yearsley}
J. M. Yearsley, D. A. Downs, J. J. Halliwell, and A. K. Hashagen,
Phys. Rev. A 84, 022109 (2011).
\bibitem{reisenberger}
M. Reisenberger and C. Rovelli, Phys. Rev. D 65, 125016 (2002).
\bibitem{marolf}
D. Marolf and C. Rovelli, Phys. Rev. D 66, 023510 (2002).
\bibitem{oppenheim}
J. Oppenheim, Ph.D. thesis, The University of British Columbia
(1999) (unpublished).
\bibitem{wootters}
D. N. Page and W. K. Wootters, Phys. Rev. D 27, 2885 (1983).
\bibitem{butkkier}
M. Buttiker and R. Landauer, Phys. Rev. Lett. 49, 1739 (1982).
\bibitem{hauge}
E. H. Hauge and J. A. Stovneng, Rev. Mod. Phys. 61, 917
(1989); M. Buttiker, Electronic Properties of Multilayers and
Low-Dimensional Semiconductor Structures, edited by J. M.
Chamberlain et al. (Plenum, New York, 1990), p. 297; R.
Landauer, Ber. Bunsenges. Phys. Chem. 95, 404 (1991); C.
R. Leavens and G. C. Aers, Scanning Tunneling Microscopy
III, edited by R. Wiesendanger and H. J. Gutherodt (Springer, ¨
Berlin, 1993), pp. 105–140; R. Landauer and T. Martin, Rev.
Mod. Phys. 66, 217 (1994).
\bibitem{oldspq}
E. O. Dias and F. Parisio, arXiv:1507.02899. (2015).
\bibitem{spq}
Eduardo O. Dias and Fernando Parisio ``Space-time-symmetric extension of nonrelativistic quantum mechanics". Phys. Rev. A 95, 032133.
\bibitem{travt}
Ricardo Ximenes, Fernando Parisio, Eduardo O. Dias ``Traversal time predictions from a space-time-symmetric quantum formalis". arXiv:1712.00397.
\bibitem{sco}
Erwin Schrodinger. ``Quantisation as a Problem of Proper Value (Part II)". Annalen der Physik(4), vol.791 1926.
\end{thebibliography}
\end{document}